\documentstyle[twoside,fleqn,espcrc2,epsf]{article}

\begin{document}

\title{Three-generation neutrino mixing and LSND dark matter neutrinos}
\author{Christian Y. Cardall and George M. Fuller
\address{Department of Physics, University of California, San Diego,
La Jolla, California 92093-0319}}
\begin{abstract}
The reported signal at the LSND experiment, when interpreted
as neutrino mixing with $\delta m^2 = 6 \;\rm{eV}^2$,
provides evidence for neutrinos with a cosmologically significant
mass. However, attempts to reconcile this interpretation of the
experiment with other hints about neutrino properties require
a (sterile) fourth neutrino and/or an ``inverted'' neutrino mass hierarchy. 
An interpretation of the LSND experiment employing 
$\delta m^2 = 0.3\; \rm{eV}^2$, with three-generation mixing and
a ``normal'' neutrino mass hierarchy, can just barely
be reconciled with the negative results of other laboratory
neutrino oscillation experiments and the positive 
hints of neutrino oscillation from the solar and
atmospheric neutrino problems. Though subject to test by 
by future experiments, such a solution allows (but does not demand)
neutrino masses relevant for dark matter.
\end{abstract}

\maketitle

\section{ACCOMODATING LSND DARK MATTER NEUTRINOS: 
	``UNNATURAL'' SCHEMES}


Based on the first year of data, 
a possible signal in the LSND experiment at Los Alamos has
been reported \cite{lsnd}. Results from the second year of data 
confirm this signal \cite{lsnd3}. An early and oft-quoted 
interpretation of this signal was that it represents 
$\overline \nu_e
\leftrightarrow \overline \nu_{\mu}$ mixing, with $\delta m^2
\approx 6 \;\rm{eV}^2$ and $\sin ^2 2\theta \approx 5 \times
10^{-3}$. The motivations for 
the selection of these mixing parameters are
twofold: (1) compatibility with the negative results of other
accelerator or reactor experiments (see Fig. 3 of Ref. \cite{lsnd});
and (2) the cosmologically interesting neutrino mass of 2.5 eV,
which fits well into cold+hot dark matter models that 
seem to satisfy many observational constraints on large-scale
structure formation \cite{primack95}.


Many attempts to reconcile the LSND result with
neutrino oscillation solutions to the solar and atmospheric 
neutrino problems have been made \cite{numod}. 
These have at least
one of two ``unnatural'' features: a fourth neutrino, and/or an
inverted neutrino mass hierarchy. 

A fourth neutrino is required when three different independent
mass differences are used, {\em e.g.}, $\delta m^2_{\rm{LSND}}
 \approx 6\; \rm{eV}^2$,  $\delta m^2_{\rm{atmos}}
 \approx 10^{-2}\; \rm{eV}^2$, and $\delta m^2_{\rm{solar}}
 \approx 10^{-5}\; \rm{eV}^2$ or $10^{-10}\; \rm{eV}^2$. 
From measurements of the width
of the $Z_0$ boson, it is known that only three neutrino species
participate in electroweak interactions \cite{zwidth}. 
Therefore, a fourth neutrino
would have to be ``sterile,'' having no standard model interactions. 
Since a $\nu_{\mu} \leftrightarrow \nu_s$ solution to the atmospheric
neutrino problem is restricted by big-bang nucleosynthesis considerations
\cite{bbn}, the sterile neutrino typically makes its appearance in a
$\nu_e \leftrightarrow \nu_s$ small-angle MSW solution to the solar
neutrino problem.

An inverted neutrino mass hierarchy is one in which the neutrino
mass eigenvalue most closely associated with the $\nu_e$ is heavier
than those associated with the $\nu_{\mu}$ or $\nu_{\tau}$, or the
mass eigenvalue associated with the $\nu_{\mu}$ is heavier than
that associated with the $\nu_{\tau}$. The requirement of an inverted
hierarchy arises from considerations of supernova $r$-process
nucleosynthesis \cite{supMSW}. 
For the typical mixing parameters associated with
the LSND experiment cited above, an MSW resonant transformation
increases the average energy of the $\nu_e$ population
in the post core-bounce supernova environment. This drives the
material around the supernova remnant proton-rich, precluding
synthesis of the $r$-process elements. This problem does not 
occur with an inverted mass hierarchy in which the $\nu_e$ is 
the heaviest neutrino. In this case, an MSW resonant transformation
increases the average energy of the $\overline \nu_e$ population,
enhancing the neutron-rich conditions required for successful
$r$-process nucleosynthesis.


Should the existence of sterile neutrinos or an inverted neutrino
mass hierarchy be designated ``unnatural''? After all, we expect
the existence of particles beyond the standard model, and we 
have no solid reason to assume that the generational mass pattern 
of the neutrinos should follow that of the corresponding charged
leptons. The relevant point is that before assuming that the
positive hints of neutrino mixing---LSND and the solar and 
atmospheric neutrino problems---point to things significantly 
different than what is expected from the standard model, 
schemes more in harmony with our experience from the standard
model should be pushed to their limits first.

\section{ACCOMODATING LSND NEUTRINOS: A ``NATURAL'' SCHEME}

We seek to construct  a ``natural'' scheme of neutrino mass
and mixing that reconciles all known hints about neutrino 
properties. Such a ``natural'' scheme would involve only 
three neutrino flavors, and would not have
an inverted neutrino mass hierarchy. We would expect such a 
solution to take advantage of the possibilities of 
three-generation mixing, rather than simply consisting of
a set of two-flavor neutrino mixings that have been ``stitched
together.'' In particular, the atmospheric neutrino problem
could involve {\em both} $\nu_{\mu} \leftrightarrow \nu_e$
and $\nu_{\mu} \leftrightarrow \nu_{\tau}$ mixing.

In using only three neutrinos, one is immediately faced with
the problem mentioned earlier, {\em i.e.} three independent
mass differences. To accomodate a neutrino mixing solution to
the solar neutrino problem, a small mass difference is required:
$\delta m^2_{\rm{solar}}
 \approx 10^{-5}\; \rm{eV}^2$ or $10^{-10}\; \rm{eV}^2$ for 
MSW and ``just-so'' vacuum oscillation solutions, respectively
\cite{krastev95}.
These scales for the mass difference cannot be altered much. In
the case of the MSW solution, a mass level crossing is the basis
of the whole effect, and so the mass difference is essentially
determined by solar parameters. The ``just so'' vacuum
oscillation solution requires that the earth-sun distance be
on the order of the neutrino oscillation length; thus the mass
difference for this solution is determined as well.

There may be a little more freedom in the neutrino mass 
differences used to explain the LSND signal and the atmospheric
neutrino problem, however. In particular, we consider the 
possibility of explaining {\em both} LSND and the atmospheric
neutrino data with a single common mass difference in the range
$\delta m^2_{\rm atmos,\;LSND} \approx 0.2-0.4$ eV$^2$. This
is an order of magnitude {\em smaller} than the mass difference
usually associated with the LSND signal, and an order of magnitude
{\em larger} than that usually associated with atmospheric neutrinos.

Is the use of a common mass difference for LSND and atmospheric 
neutrinos feasible? Fig. 3 of Ref. \cite{lsnd} shows that 
$\delta m^2_{\rm atmos,\;LSND} \approx 0.2-0.4$ eV$^2$ is
compatible with the negative results of other accelerator and
reactor experiments. However, the claimed zenith-angle dependance
of the atmospheric neutrino data seem to imply a smaller mass
difference, $\delta m^2_{\rm atmos} \approx 10^{-2}$ eV$^2$, 
with a 90\% C.L. upper limit of about 0.1 eV$^2$
\cite{fukuda94}. We note that 
the statistical significance of this fit has been questioned
\cite{saltzberg95}. 
According to another analysis, of the Kamiokande multi-GeV data,
$\delta m^2_{\rm atmos} \ge 0.25$ eV$^2$ is excluded at 
90\% C.L., and $\delta m^2_{\rm atmos} \ge 0.47$ eV$^2$
is excluded at 95\% C.L. \cite{yasuda96}.
Thus $\delta m^2_{\rm atmos,\;LSND} 
\approx 0.3$ eV$^2$ would be allowed at 95\% C.L.


Next we turn to a discussion of the mixing angles. 
The neutrino flavor eigenstates, $\nu_{\alpha}$, 
can be written as 
a linear combination of mass eigenstates,
$\nu_i$:
\begin{equation}
\nu_{\alpha} = \sum_i U_{\alpha i} \nu_i,
\end{equation}
where $U_{\alpha i}$ are elements of a unitary mixing
matrix $U$. We take $U$ to be a standard parametrization
of the Cabbibo-Kobayashi-Maskawa (CKM) matrix involving
three mixing angles and a CP violating phase \cite{revpart}.

The mass
differences indicated above satisfy $\delta m^2_{\rm solar}
\ll \delta m^2_{\rm atmos,\;LSND}$. A useful approximation 
in this case is the ``one mass scale dominance'' (OMSD) limit,
in which we neglect $\delta m^2_{\rm solar}$ relative to 
$\delta m^2_{\rm atmos,\;LSND}$. The
discussion is considerably simplified in this limit, as
the results from two-flavor interpretations of neutrino
mixing experiments can be directly converted to the 
three-neutrino OMSD interpretation \cite{fogli95}. 
For two-neutrino
vacuum oscillations, the survival probability $P$ is 
given by
\begin{equation}
P = 1- \sin ^2 2\theta \sin ^2 \left(1.27 {\delta m^2 L
\over E} \right),
\end{equation}
where $L$ is the path length (in km) of a neutrino initially
in a flavor eigenstate at $L=0$, $E$ is the neutrino energy in
GeV, and $\delta m^2$ ($=\delta m^2_{\rm atmos,\;LSND}$) 
is in eV. The two-flavor mixing angle is
$\theta$. In the OMSD limit, we have the following correspondence
between the two-flavor mixing angle and the elements of the
three-flavor mixing matrix:
\begin{eqnarray}
\sin^2 2\theta \Leftrightarrow &4 |U_{\alpha 3}|^2 |U_{\beta 3}|^2 \; \;
  &(\rm{app.}),   \label{corres}\\
\sin^2 2\theta \Leftrightarrow &4 |U_{\alpha 3}|^2 
(1-|U_{\alpha 3}|^2) \; \;
  &(\rm{disapp.});
\end{eqnarray}
for appearance and disappearance experiments, respectively.
For the parametrization of $U$ in Ref. \cite{revpart}, we have
\begin{eqnarray}
|U_{e3}|^2 = &&\sin^2 \theta_{13}, \\
   |U_{\mu 3}|^2 = &&\cos^2 \theta_{13} 
\sin^2 \theta_{23}, \\
   |U_{\tau 3}|^2 = &&\cos^2 \theta_{13} \cos^2 \theta_{23}. 
\end{eqnarray}
Notice that the mixing angle $\theta_{12}$ and the CP-violating
phase do not appear in the oscillation probability. Thus
$\delta m^2$ and the mixing angles $\theta_{13}$ and $\theta_{23}$
are the only parameters needed to describe three-flavor mixing
involving the larger mass difference ($\delta m^2_{\rm atmos,\;LSND}$)
in the OMSD case.

Fig. 1 shows the allowed ranges of the mixing angles $\theta_{13}$
and $\theta_{23}$. Each panel corresponds to a different value of
$\delta m^2_{\rm atmos,\;LSND}$. This figure shows that a small 
area of the $\theta_{13}$ and $\theta_{23}$ plane is available for 
$\delta m^2_{\rm atmos,\;LSND} \approx 0.3-0.4 \;{\rm eV}^2$
which explains the LSND signal, provides a solution to the atmospheric
neutrino problem, and does not conflict with the negative results of
other accelerator and reactor experiments. For small values of 
$\theta_{13}$---which are indicated by the solution in Fig. 1---the
small angle MSW solution to the solar neutrino problem is essentially
unaffected \cite{fogli94}. 

Thus it appears that there is an essentially unique solution
that can simultaneously explain the LSND signal and solve the
solar and atmospheric neutrino problems, all without introducing
sterile neutrinos. We have shown elsewhere that
existing limits on two-flavor mixing parameters 
based on supernova $r$-process
nucleosynthesis can be applied to the three-flavor mixing case in
the OMSD limit \cite{cardall96}. 
We conclude that the above solution 
has a small enough mass difference that supernova $r$-process
production is not adversely affected \cite{supMSW}.

Cold+hot dark matter models,
with about 20\% of the dark matter comprised of two species of $\sim$2.5 eV
neutrinos, are reported to fit large-scale structure data well
\cite{primack95}.
Choosing $\delta m^2_{\rm LSND} \approx 0.3-0.4 \;{\rm eV}^2$ no longer
directly implies the existence of neutrinos with this mass. Since neutrino
oscillations are only sensitive to neutrino mass {\em differences}, however,
all of the neutrino masses can be offset from zero to provide a 
neutrino mass sum $\sim 5$ eV. In such a scheme the
neutrino masses would be nearly degenerate, a possibility considered
in Refs. \cite{degen}.\footnote{Note 
that with all three neutrino masses offset
from zero, the OMSD limit (which applies to mass {\em differences}) can
still apply, even if the masses themselves are ``nearly degenerate.''}

It should be noted that the putative solution in Fig. 1
is fragile. We have already mentioned
that the mass difference 
$\delta m^2_{\rm atmos,\;LSND} \approx 0.3-0.4 \;{\rm eV}^2$
is allowed at 95\% C.L., but not at 90\% C.L. A similar situation
exists for the mixing angles: the contours in Fig. 1 (except the
atmospheric neutrino solution and LSND contours) are 95\% C.L. contours.
At 90\% C.L., the LSND detection band shrinks and the 
exclusion region from other accelerator and reactor experiments
expands, and the solution in Fig. 1 virtually disappears.

We have sought a ``natural'' solution, but in the end it still has
what might be considered some ``unnatural'' features. 
For example, we have taken $\delta m^2_{\nu_e \nu_{\tau}} \approx 
\delta m^2_{\nu_{\mu} \nu_{\tau}} \gg \delta m^2_{\nu_e \nu_{\mu}}$,
in analogy with the hierarchy of mass differences 
of the charged leptons.
For the charged leptons, this hierarchy of 
mass {\em differences} arises naturally
from the hierarchy of the {\em masses themselves}, 
i.e. $m^2_e \ll m^2_{\mu} 
\ll m^2_{\tau}$.
However, the offset of neutrino masses from zero 
required to make the neutrino
masses sum to about 5 eV for cold+hot dark matter 
models causes the absolute 
masses to have roughly similar magnitudes, in 
contrast to the masses of the 
charged leptons.

Another unnatural feature of this putative ``natural'' 
solution is that 
several of the off-diagonal 
elements of the mixing matrix $U$ have relatively 
large magnitudes. 
This is in contrast 

\epsfysize=7cm \epsfbox{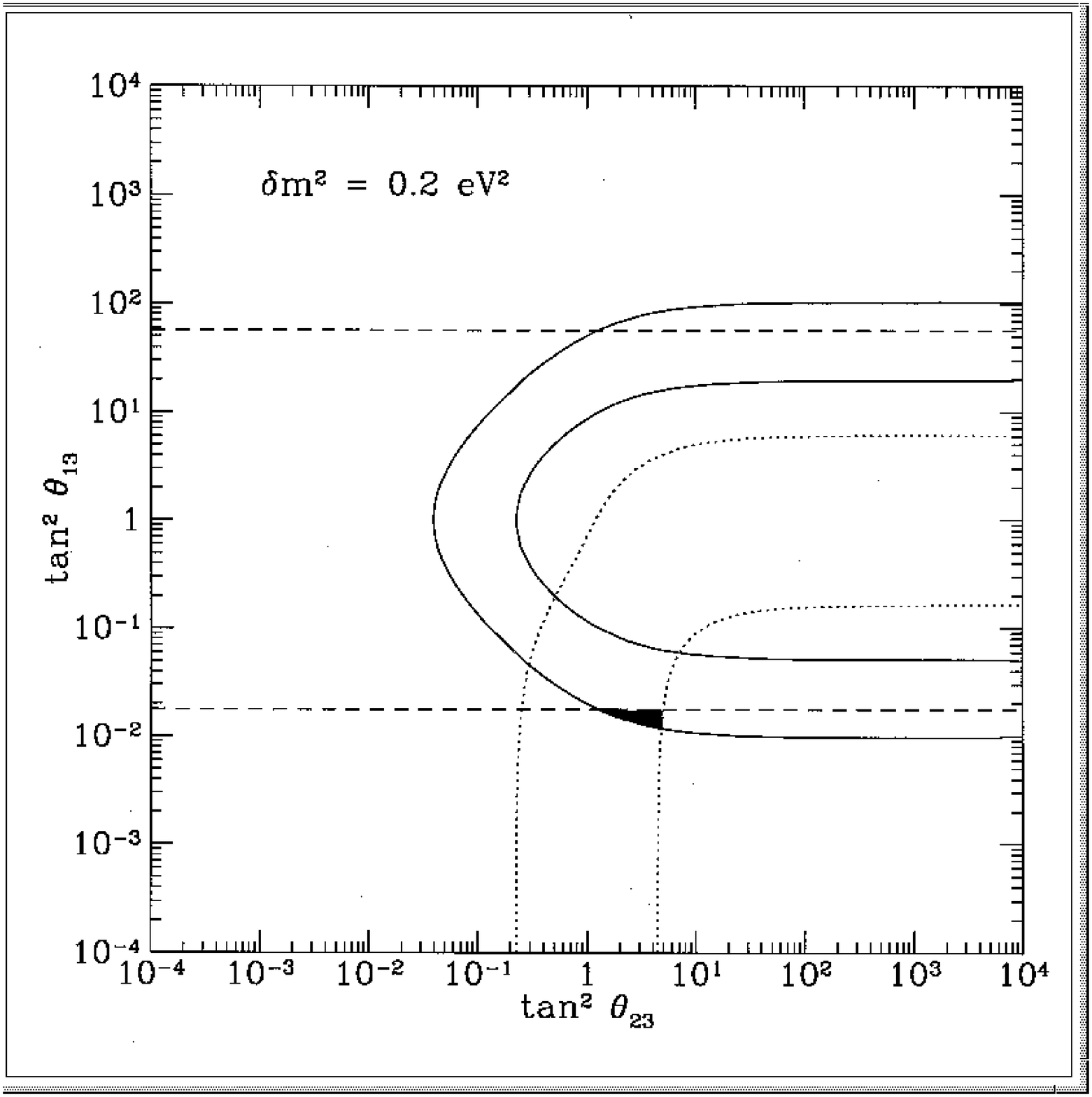}

\epsfysize=7cm \epsfbox{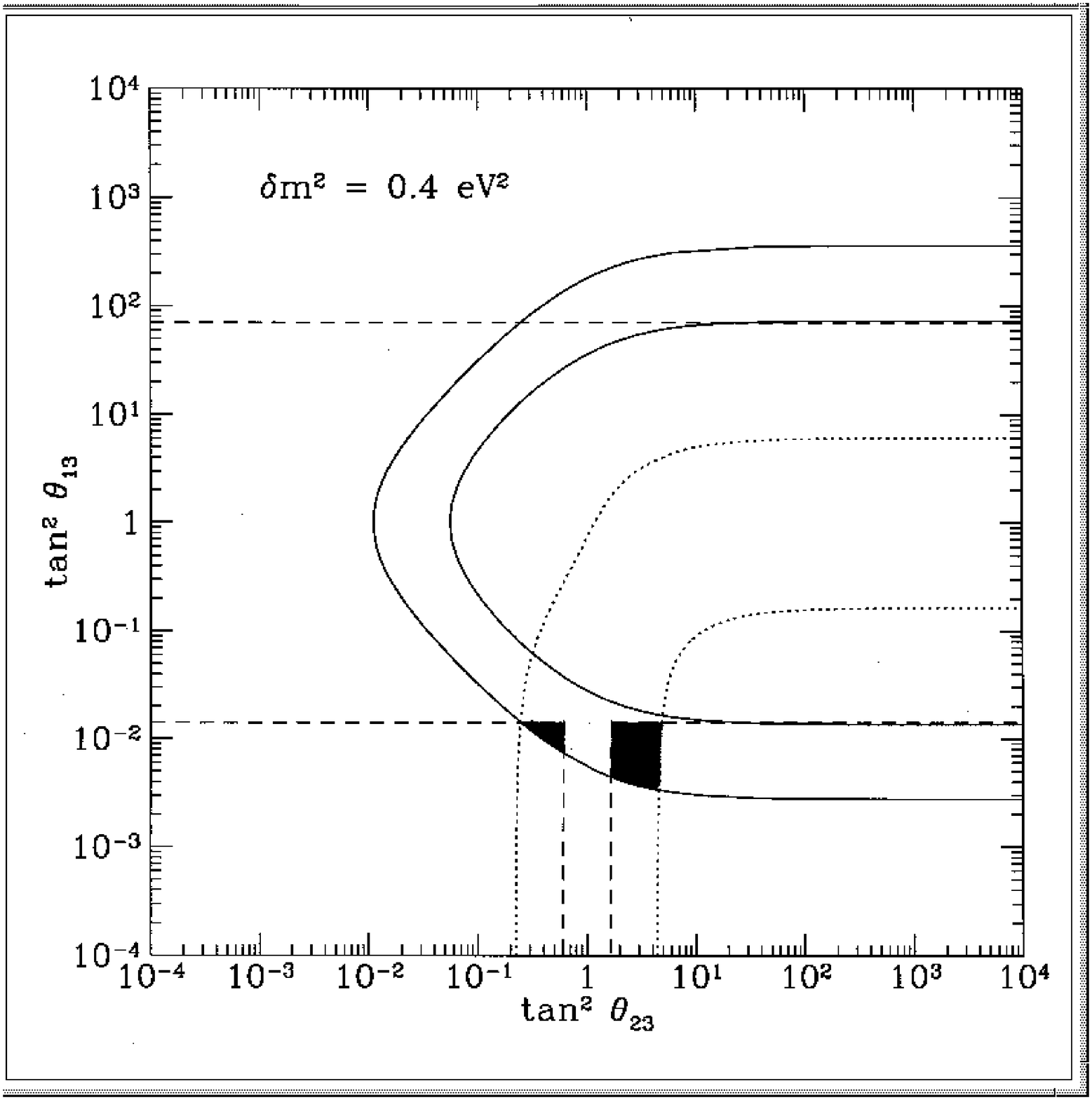}

\noindent{ 
\hbox to 16cm{Figure 1. Allowed regions of the mixing 
angles $\theta_{13}$ and $\theta_{23}$,
for four values of the dominant mass difference.} 
\hbox to 16cm{The region between the
solid lines is the ``99\% likelihood'' detection region of LSND \cite{lsnd3}. 
The region inside} 
\hbox to 16cm{the long dashed lines is excluded by accelerator/reactor 95\%
C.L. limits \cite{fogli95}. The region enclosed by }
\hbox to 16cm{the short dashed lines 
represents a solution to the sub-GeV atmospheric neutrino data \cite{atmos}.}
}
\vskip 0.2cm
\noindent to the quark mixing case \cite{revpart}. It has
been recognized previously that a three-neutrino 
oscillation explanation of
the LSND experiment requires this unusual feature 
\cite{bilenky95}. 
Large off-diagonal
terms will generally be present whenever
 oscillation  probabilities are
large, and neutrino oscillation explanations of 
the atmospheric neutrino 
anomaly and the LSND data both invoke 
relatively large oscillation
probabilities.

\epsfysize=7cm \epsfbox{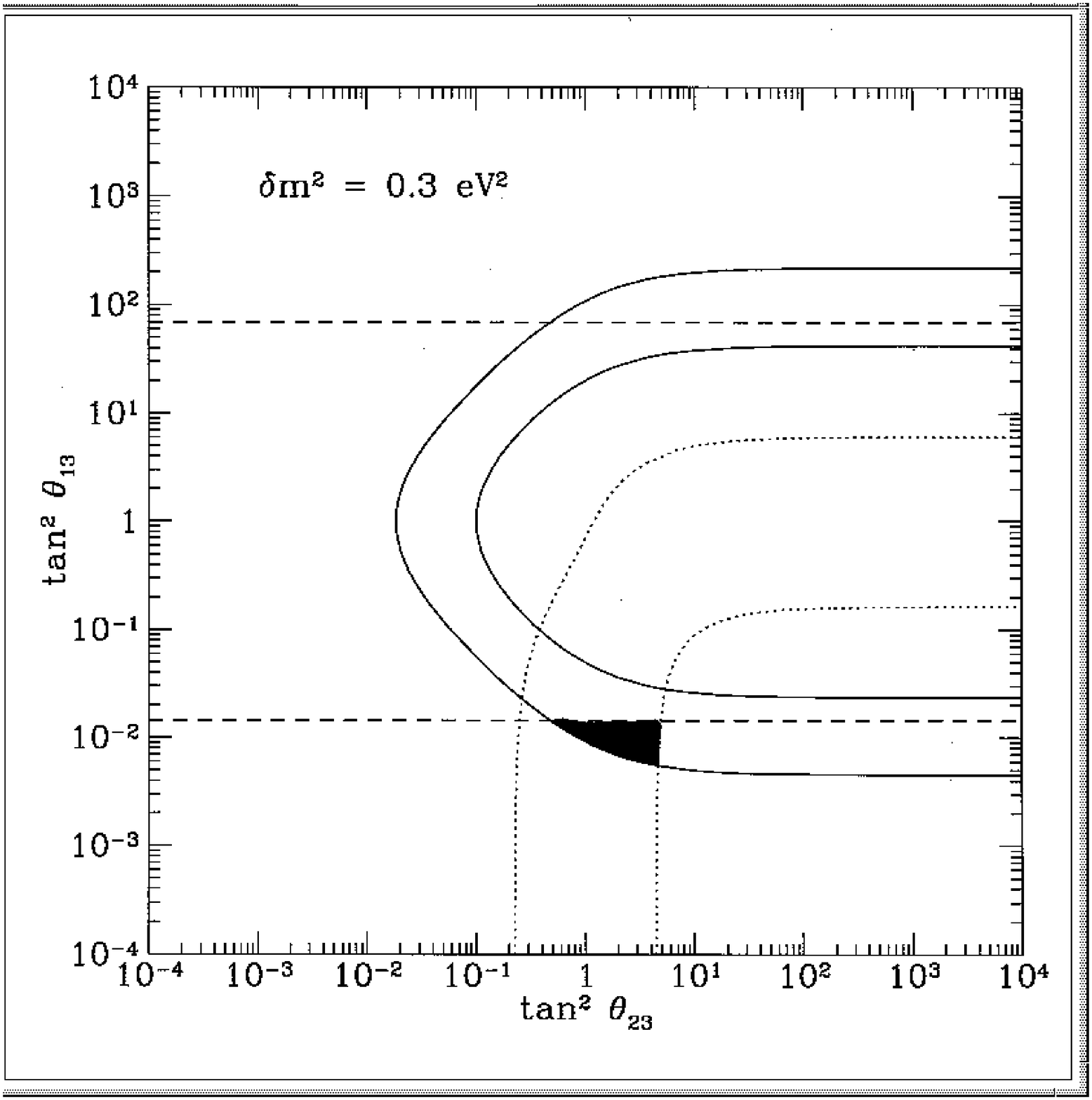}

\epsfysize=7cm \epsfbox{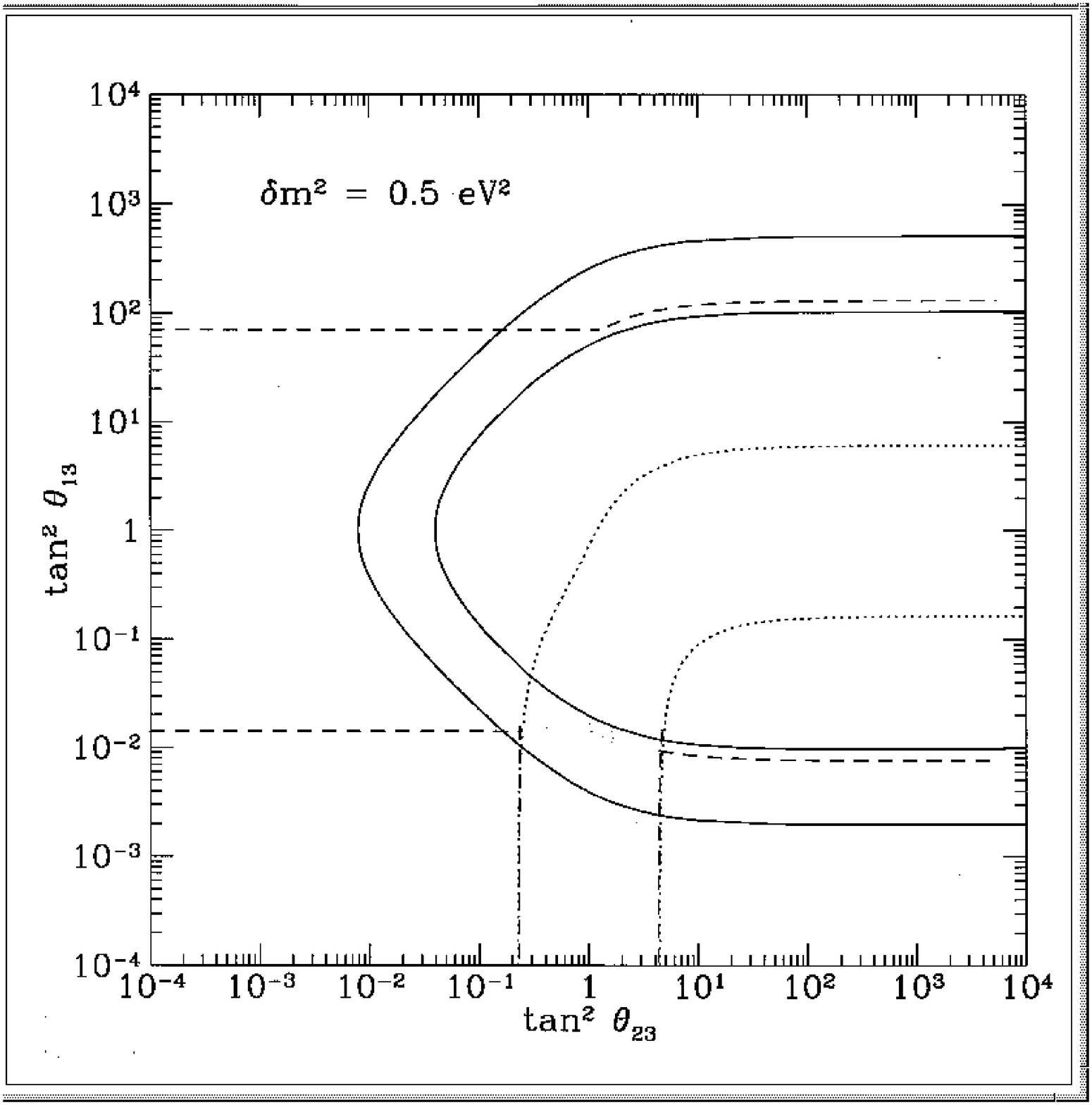}
\vskip 2.3cm

\section{FUTURE TESTS}

Fortunately, the prospects are good for testing the last
remaining three-neutrino solution that explains the LSND signal
and solves the solar and atmospheric neutrino problems. 
The LSND experiment continues to run. A comparable experiment,
KARMEN, is receiving an upgrade that within a few years should
provide either a confirmation or refutation of the LSND signal
\cite{kleinfeller96}.
Super-Kamiokande will provide much better statistics on the 
claimed zenith-angle dependance of the atmospheric neutrino data
\cite{superkam}.
Super-Kamiokande and SNO may also be able to distinguish solar
neutrino solutions involving sterile neutrinos from solutions
involving only active neutrino flavors \cite{nuexp}.
A null result at the San Onofre reactor experiment 
\cite{sanonofre} would
not rule out the ``natural'' solution proposed here;
on the other hand, a detection of neutrino oscillations
at San Onofre could not be accounted for by our solution.
However, proposed long-baseline accelerator 
experiments will probe the entire region of 
parameter space favored by 
the atmospheric neutrino sub-GeV data 
\cite{fogli95,michael95} and could thus
provide the crucial test.

Supported by grants NSF PHY95-03384 and NASA NAG5-3062
at UCSD.

\end{document}